\documentclass[reprint,amsmath,amssymb,aps,]{revtex4-2}

\usepackage{graphicx}
\usepackage{bm}
\usepackage{xcolor}
\usepackage[colorlinks=true, urlcolor=blue, linkcolor=blue, citecolor=green]{hyperref}

\usepackage[normalem]{ulem}

\begin{document}

\newcommand{\sg}[2]{\textcolor{red}{\sout{#1}}\textcolor{red}{#2}}

\preprint{APS/123-QED}

\title{Investigation of the ensemble of maximal center gauge}

\author{Zeinab Dehghan}
\email{zeinab.dehghan@ut.ac.ir}
\affiliation{Department of Physics, University of Tehran, Tehran 14395-547, Iran}
\affiliation{School of Physics, Institute for Research in Fundamental Sciences (IPM),
P. O. Box 19395-5531, Tehran, Iran}

\author{Rudolf Golubich}%
\email{rudolf.golubich@gmail.com}
\affiliation{ 
Atominstitut, Technische Universität Wien
}%

\author{Roman Höllwieser}
\email{hoellwieser@uni-wuppertal.de}
\affiliation{%
Department of Physics, Bergische Universität Wuppertal
}%

\author{Manfried Faber}%
\email{faber@kph.tuwien.ac.at}
\affiliation{ 
Atominstitut, Technische Universität Wien
}%
 
\date{\today}

\begin{abstract}
Maximal Center Gauge (MCG) aims to detect center vortices by maximizing a gauge functional and then projecting onto the center elements of the respective group. The requirement for unrestricted maximization of the gauge functional has proven to be untenable because it was shown that it leads to an underestimation of the string tension. To counter this problem,  the ensemble of local gauge maxima is investigated and it is found that the unrestricted maximization can be replaced by a maximization restricted to the Gaussian distributed part of the ensemble. Such restricted maximization weakens the problem of an underestimated string tension.
\end{abstract}

\maketitle


\section{Introduction}
Important nonperturbative properties of the QCD vacuum are confinement~\cite{greensite:2011zz} and dynamical breaking of chiral symmetry~\cite{faber:2017alm}. They show that the QCD vacuum is highly non-trivial and filled with quantum fluctuations. The amount and type of these quantum fluctuations depend on the temperature. The most interesting quantum fluctuations that determine the properties in the hadronic phase are abelian monopoles~\cite{Hooft81,Kronfeld:1987ri,Bornyakov_2022} and center vortices~\cite{thooft:1979rtg,deldebbio:1996mh}. Abelian monopoles are defined by a charge in a U(1) subgroup of SU($N$). They move along paths that percolate through the four-dimensional space-time. Center vortices are closed color-magnetic flux tubes, quantised to center elements of the gauge group and form two-dimensional surfaces percolating the vacuum. Projecting the color of the flux tubes into a U(1) subgroup results in trajectories for abelian monopoles~\cite{Chernodub:2001ht,Suganuma:2021mcv}. The density of these topological excitations determines the strength of the force between color charges, the QCD string tension~\cite{Shibata:2019bke,Junior:2021vpd}. Successful methods for identifying these topological objects are special gauges, which are determined by maximizing corresponding gauge functionals. The maximal abelian gauge for magnetic monopoles and the maximal center gauge (MCG) for vortices have proven to be particularly successful. We examine maximal center gauge for the SU(2) gauge theory formulated with the Wilson action~(\ref{WilsonAction}).

Local maxima of a gauge functional can only be searched for numerically. In the maximal Abelian gauge, the highest gauge functionals provide the best description of the QCD vacuum and justify the maximization prescription. Local maxima with high values of the gauge functional could also be found for MCG, which predict the correct string tension. However, a more precise search led to maxima~\cite{kovacs:1998xm,bornyakov_2001} which provide a string tension that is much too low. Furthermore, it was shown in ~\cite{dehghan_2023} that with series of random gauge copies with broken local Z(2) gauge symmetry due to restricting to links with a positive trace, gauge configurations with a high value of the gauge functional and a low string tension can easily be generated. These observations strongly questioned the maximal center gauge and thus the vortex mechanism as an explanation of confinement. The aim of this work is to show that the error lies in the requirement of unrestricted maximization of the gauge functional and that it is not the vortex mechanism itself that is in question. Refs.~\cite{rudolfgolubich93027173,golubich2021,rudolfgolubich68347507} also point in this direction, in which non-trivial center regions are determined using a complex computational procedure in order to guide the gauge fixing.

In this article we explore ensembles of gauge copies for maximal center gauge to see whether statistical analysis allows for a less resource intensive extraction of the correct string tension as in ~\cite{rudolfgolubich93027173,golubich2021,rudolfgolubich68347507}.

After describing the formalism of maximal center gauge in the second section, we explain in the third section the problem of the global maximization prescription of the gauge functional of maximal center gauge. On the way to the modification of this prescription we examine in the forth section the distribution of local maxima of the gauge functional. We observe that in a wide range of the coupling parameter $\beta$ their densities come very close to Gaussian distributions. In the upper tip of these distributions we find agreement of the center projected string tensions with the string tensions extracted from unprojected fields. At the upper boundary of the $\beta$-window, we find that strong deviations from Gaussians are first indications for a failing absolute maximization of the gauge functional. In the further part of the work we try to extend the $\beta$-window to higher values. First, we remove these deviations by symmetrisation of the distribution. Another indication is given by the behavior of the slope of the quark-antiquark potentials of center projected fields. These potentials are almost linear, as we find in section five. This is an observation that has already been made in the determination of string tensions from Creutz ratios~\cite{hollwieser:2015koa}. In general, the slope of the potential increases slightly with the size of Wilson loops. For increasing asymmetry of the gauge functional distributions, this behavior may invert to its opposite, especially for gauge configurations with particularly high values of the gauge functional. We find then a decrease in string tension with the size of the Wilson loops. This may indicate that the detected P-vortices break apart and do no longer percolate, as was already assumed in \cite{golubich2022improvement}. By removing these vortex detection failures in section six and symmetrisation of the distributions we can successfully increase the region where the right tip of the Gaussian distributions reproduces the asymptotic string tension. In section seven we draw from these results the conclusion, how to modify the prescription for the maximization procedure of MCG.

\section{Direct maximal center gauge}
We work on a $24^4$ lattice with periodic boundary conditions and investigate field configurations at the inverse coupling constants $\beta={4}/{g^2}\in\{2.3, 2.4, 2.5, 2.6, 2.7\}$ of the Wilson action for an SU(2) gauge theory
\begin{equation}\label{WilsonAction}
S_\text{gluons}=\beta \sum_{x,\mu<\nu}\left(1-\frac{1}{2}\mathfrak{Re}
\text{Tr}(U_{\mu\nu}(x))\right).
\end{equation}
The path ordered product of links $U_\mu(x)$ delimiting a plaquette in the $\mu\nu$ plane defines the plaque variable $U_{\mu\nu}(x)$.

In direct maximal center gauge (DMCG), the link variables are transformed by a SU(2) gauge transformation $g(x)$ shifting them as close as possible to center elements by maximizing the gauge functional
\begin{equation}\label{GaugeFunctR}
R_\text{MCG}=\sum_x\sum_\mu\mid\text{Tr}[{}^gU_\mu(x)]\mid^2,
\end{equation}
where the transformed link variables are ${}^gU_\mu(x)=g(x+\hat\mu)\,U_\mu(x)\,g^\dagger(x)$. These are then projected onto the nearest center element
\begin{equation}\label{CentrProj}
{}^gU_\mu(x)\rightarrow Z_\mu(x)\equiv \mathrm{sign Tr}[{}^gU_\mu(x)].
\end{equation}
A single negative link variable $Z_\mu(x)=-1$ in a field of trivial link variables $Z_\mu=+1$ forces the values of the six attached plaquettes to $Z_{\mu\nu}(x)=-1$. On the dual lattice these six so-called P-plaquettes form the surface of a cube. It turns out that in the confinement phase, a large number of negative links are arranged in such a way that the dual P-plaquettes form extended closed surfaces: vortices percolating through the whole space-time lattice. The success of the vortex model of confinement is founded in the fact that the Abelian Stoke's law applies to Abelian and therefore also to center vortices. The sign of Wilson loops is flipped by the piercing of thin projected and also of thick vortices. A high density of random vortices leads to a fast decrease of the expectation values of Wilson loops with the size of the loops and thus to large string tensions.

In the unprojected field configurations, vortices are thick magnetic flux tubes. MCG assumes that despite the strong limitation of the field degrees of freedom in the projection procedure from SU(2) to Z(2), MCG is able to detect the shape of thick vortices and therefore to preserve the important long-range properties of the vacuum configurations.

\section{The problem with the maximization}
The gauge functional~(\ref{GaugeFunctR}) has a huge number of local maxima. Determining the global maximum of the gauge functional~(\ref{GaugeFunctR}) is an NP-hard problem, therefore only local maxima can be approached by the conjugate gradient method and the global maximum can be approximated step by step only. 
As Engelhardt and Reinhardt~\cite{Engelhardt:1999xw} have concluded from analytical considerations, a trivial maximization prescription cannot be expected to reproduce the correct physics. A reduction of the string tension with increased values of the gauge functional was numerically confirmed by Kovacs and Tomboulis~\cite{kovacs:1998xm} and Bornyakov, Komarov and Polikarpov~\cite{bornyakov_2001}: the largest maxima of the functional  underestimate the density of vortices and therefore also the string tension. A possible reason for this is that in the large volume of thick smooth vortices the center charge is spread over multiple links causing them to be projected onto the trivial center. The more parts of the thick vortices the P-vortices miss, the higher are the values of the gauge functional. With rising beta values, the field configurations generated by the Monte Carlo method and the Metropolis algorithm become too smooth and it is increasingly difficult to describe the shape of thick vortices by sudden changes in the link variables with values of $\pm1$.

However, the investigations so far show that the gauge functional~(\ref{GaugeFunctR}) of MCG is in principle able to predict the string tension. Here we would like to mention, (i) the successful predictions by DMCG~\cite{DelDebbio:1998luz,hollwieser:2008tq} where only a few local maxima were examined, (ii) the pre-selection of gauge copies with the help of the eigenfunctions of the Laplaceoperator~\cite{Faber_2001,Faber_2002} which lead to good results and (iii) averages over the ensemble of local maxima of the gauge functional~\cite{dehghan_2023} which did not suffer from extreme values of $R_\text{MCG}$.

The previous prescription for MCG striving for maximization was based on the assumption that the global maximum corresponds to the most physical gauge. This objective has clearly been missed due to the nearly linear decrease of the string tension~\cite{dehghan_2023} with increasing value of the gauge functional.

The above enumerated indications encourage us to examine how the information about the value of string tension is encoded in the properties of the ensembles of the local maxima. To accomplish this, we determine the distribution of the ensemble of local maxima and the relation between the values of the gauge functional and the string tension. Throughout these investigations, we hope to find hints for which cases the maximization of the gauge functional leads to incorrect predictions of string tension. Finally, this should lead us to a modification of the MCG prescription for the vortex detection.

\begin{figure}[h!]
\centering
a)\hspace{-10mm}\includegraphics[scale=0.838]{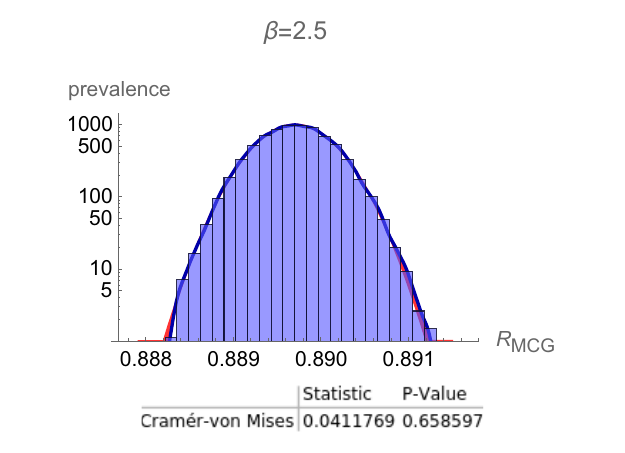}\\
b)\hspace{-10mm}\includegraphics[scale=0.838]{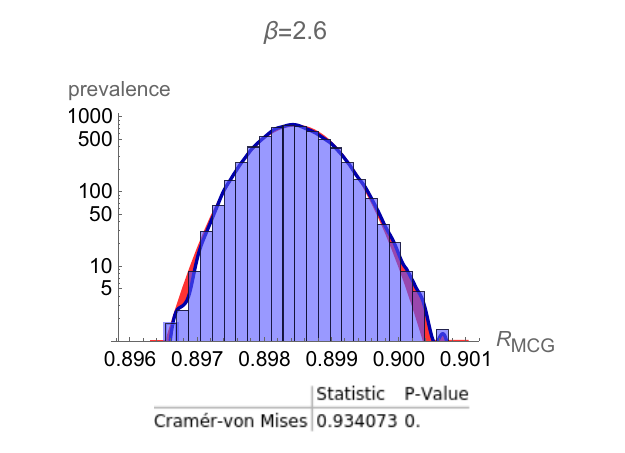}\\
c)\hspace{-10mm}\includegraphics[scale=0.838]{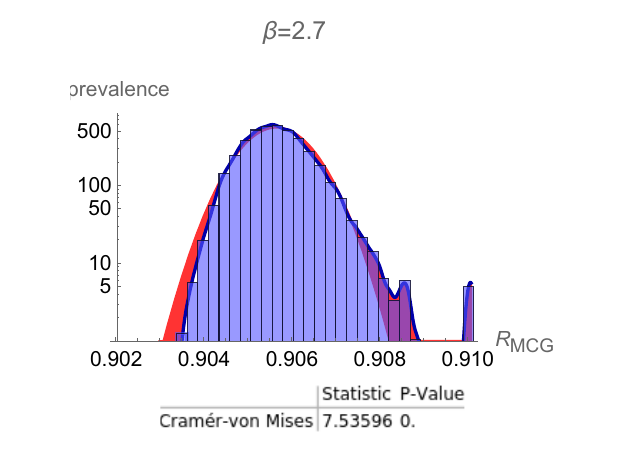}
\caption{The distributions of the ensembles of 20\,000 local gauge maxima are depicted for $\beta=2.5$, $\beta=2.6$ and $\beta=2.7$ in logarithmic plots. A smooth kernel is shown as blue solid line with the histogram of the data placed in the background. Deviations from the normal distribution are colored red. Results of distribution tests are shown below each distribution. One can clearly see increasing deviations from Gaussians and with $\beta$ increasing skewness.}
\label{fig:DistriDistortion}
\end{figure}
\section{Distributions of local maxima}
\begin{figure}[h!]
\includegraphics[width=\linewidth]{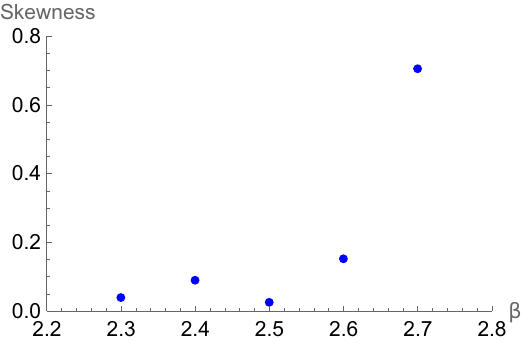}
\caption{The skewness of the distributions of the local gauge maxima is depicted for different values of $\beta$.}
\label{fig:DistriSkewness}
\end{figure}

The distribution of the gauge functional values~(\ref{GaugeFunctR}) at the local maxima resembles a Gaussian, see Fig.~\ref{fig:DistriDistortion}.
We have investigated these distributions for $\beta \in \{2.3, 2.4, 2.5, 2.6, 2.7\}$ by Monte-Carlo runs with 20 random starts, 3000 initial sweeps and 10 configurations with a distance of 1000 sweeps. For each of these 200 configurations we performed 100 unbiased random gauge copies leading to 20\,000 values of the gauge functional for each $\beta$. In Fig.~\ref{fig:DistriDistortion} 
we depict logarithmic plots of these distributions for $\beta=2.5, 2.6$ and $2.7$ and bin widths of one-third of the standard deviation. A continuous replacement for histograms, a smooth kernel distribution, is shown by the full blue lines. Their differences to Gaussian fits are filled by red areas. Below the distributions we display the values of the Cram$\acute e$r–von Mises criteria for judging the goodness of the fit to the normal distributions. p-values below 0.05 indicate statistically significant deviations from normal distributions.
Until a specific value of $\beta$ the ensemble of gauge copies is well described by a normal distribution. A measure for the deviations from symmetric distributions is the skewness, the ratio of the third central moment to the 3/2-power of the second central moment. With increasing $\beta$ the skewness of the distribution increases as can be seen in Fig.~\ref{fig:DistriSkewness}.
This lets us suspect that a partial trivialization of the lattice may occur during gauge fixing and projection. This effect increases with $\beta$ and thus an increasing proportion of gauge copies of the ensemble is shifted to larger values of $R_\text{MCG}$ as is hinted in Fig.~\ref{fig:DistriDistortion}b and can be nicely seen in Fig.~\ref{fig:DistriDistortion}c. As we will show in the next sections, this corresponds to a decrease in the string tension extracted from the expectation values of Wilson loops.

Fig.~\ref{fig:DistriDistortion} shows that deviations from the normal distribution increase with increasing value of $\beta$. We know already that for the local maxima with the highest values of the gauge functional~(\ref{GaugeFunctR}), the vortex detection fails due to the above mentioned smoothness of the lattice. This suggests to remove these misguided detections by symmetrizing the distributions.

\begin{figure}[ht!]
\includegraphics[width=.9\linewidth]{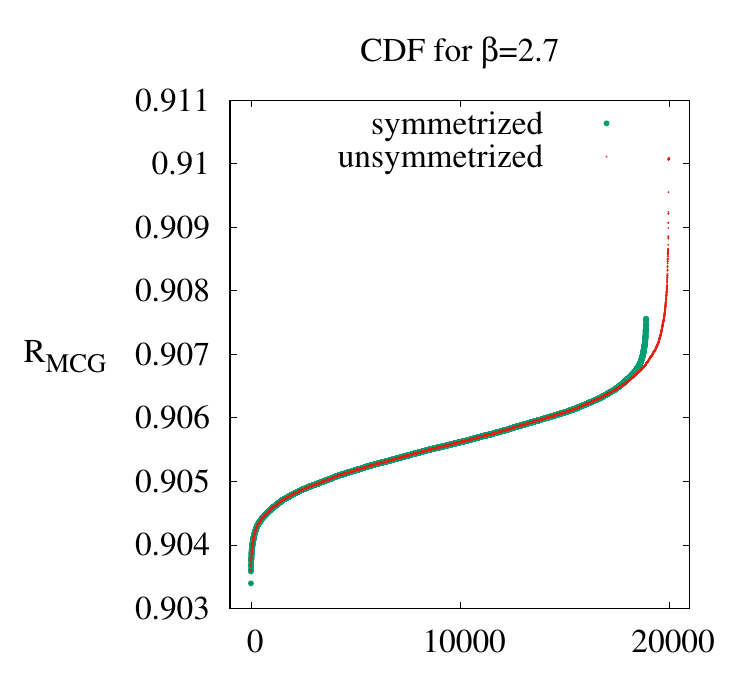}
\caption{Cumulative distribution functions (CDF) of local maxima of gauge functional~(\ref{GaugeFunctR}) for $\beta=2.7$ before and after symmetrisation of the distribution.}
\label{fig:CDF}
\end{figure}

The distributions for $\beta=2.3, 2.4$ and $2.5$ are not affected by asymmetries and should remain unchanged, while $\beta=2.6$ should only be slightly modified and $\beta=2.7$ should be strongly changed. The cumulative distribution function (CDF) of local maxima lends itself well to symmetrisation. This makes it possible to thin out the high ranges of $R_\mathrm{MCG}$-values and leave the remaining ranges unchanged. Fig.~\ref{fig:CDF} shows the CDFs before and after symmetrisation for $\beta=2.7$.

\section{String tensions extracted from center projected Wilson loops}
\begin{figure}[h!]
\includegraphics[width=.9\linewidth]{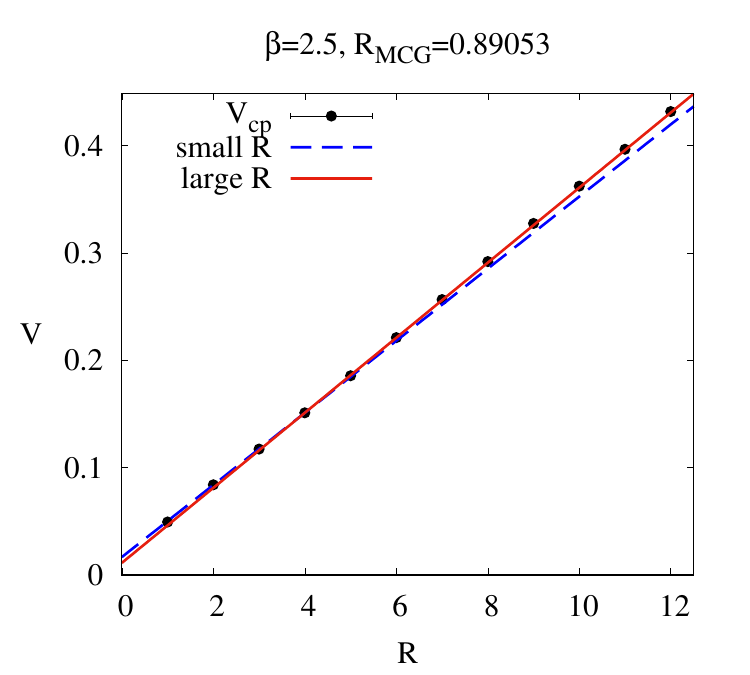}
\caption{Potential $V_\mathrm{cp}(R)$ of center projected Wilson loops as functions of their spatial extent $R$ extracted at $\beta=2.5$ from 200 configurations symmetrically arranged around $R_\text{MCG}=0.89053$. The linear approximations for small and large $R$ demonstrate a slight increase in slope due to short range fluctuations which do not influence the long range behavior.}
\label{fig:UsualV}
\end{figure}

From expectation values of center projected Wilson loops $W_\mathrm{cp}(R,T)$ we extract the potential
\begin{equation}\label{VfromW}
V_\mathrm{cp}(R)=\lim_{T\to\infty}\frac{1}{T}\ln W_\mathrm{cp}(R,T)
\end{equation}
and a slope $\sigma_\mathrm{cp}$. A characteristic $R$ dependence of $V_\mathrm{cp}$ is shown in Fig.~\ref{fig:UsualV}. 
The same nearly linear behavior was also observed for Creutz ratios, see Ref.~\cite{hollwieser:2015koa}. It is explained by center vortices randomly piercing the minimal areas of Wilson loops. We expect this behavior for vortices percolating the lattice. Short-range fluctuations of vortex surfaces~\cite{bertle:1999tw} lead to correlated pairs of piercings for large loops and do not influence their expectation values, but they increase the number of single piercings for
small loops and decrease therefore the slope of the potential at short distances. This explains the slight increase of the slope with R for the characteristic example in Fig.~\ref{fig:UsualV} from $\sigma_\mathrm{cp}=0.0336\pm3$ to $\sigma_\mathrm{cp}=0.0350\pm3$.

\begin{figure}[h!]
\includegraphics[width=\linewidth]{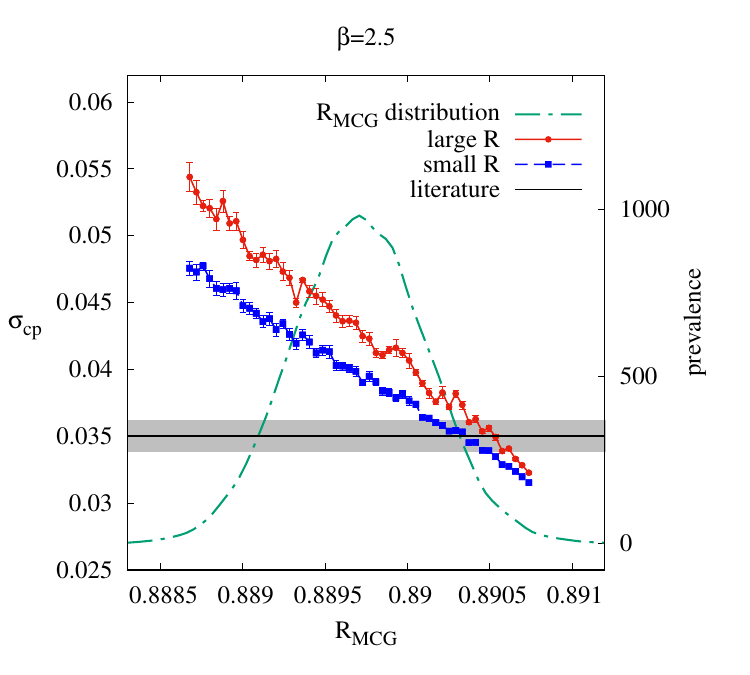}
\caption{Comparison of the slopes $\sigma_\mathrm{cp}$ of the potential $V_\mathrm{cp}(R)$ with the literature value~\cite{Bali1994} of the string tension. The expectation values of the center projected Wilson loops were averaged over 200 local maxima around the indicated $R_\mathrm{MCG}$ values. For comparison the nearly Gaussian distribution of the functional is shown as a dash-dotted line. The most interesting region is the intersection of the red data with the literature value. In the right tip of the distribution, we observe the best agreement of $\sigma_\mathrm{cp}$ with the asymptotic string tension.}
\label{fig:bothsigmasRR25}
\end{figure}

For Creutz ratios, we found~\cite{dehghan_2023} that the $\sigma_\mathrm{cp}$ values are approximately linear functions of the local maxima of $R_\mathrm{MCG}$. A corresponding correlation for the slope $\sigma_\mathrm{cp}$ of the potential $V_\mathrm{cp}(R)$ is shown in Fig.~\ref{fig:bothsigmasRR25} 
 at small $R$ values in blue and large $R$ in red color. The string tensions extracted from unprojected configurations~\cite{Bali1994} is shown by a horizontal black line. The slope of the potential is generally greater for large distances than for short distances as explained in the discussion of  Fig.~\ref{fig:UsualV}. We used 200 configurations around the indicated value of $R_\mathrm{MCG}$ for the determination of the potential. Due to the fluctuations of the vortex shapes a sample of 200 fields out of the 20\,000 is obviously still not enough to get a smooth curve. The larger variations in the slope of the potential for ``large $R$'' than for ``small $R$'' are due to the statistical errors that increase with the size of the Wilson loops. On average, an almost linear relationship between the gauge functional and the string tension is observed. In order to be able to classify the $R_\mathrm{MCG}$ values, we show their normalized density distribution for comparison. With this diagram, we want to focus on the important region of the distribution of $R_\mathrm{MCG}$ values. We can clearly see that it is the right tip of the distribution around $R_\mathrm{MCG}=0.8905$ where the slopes of the potentials at large distances approaches the literature value~\cite{Bali1994} of the string tension. There the density of local maxima of the gauge functional has already decreased by a factor of about 10.

For the lowest $\beta$ values, $V_\mathrm{cp}(R)$ can only be studied for smaller spatial extents of Wilson loops, so only the asymptotic string tension can be determined. In the $2.4\le\beta\le2.6$ range, we find similar behavior as for $\beta=2.5$.

\begin{figure}[h!]
\includegraphics[width=\linewidth]{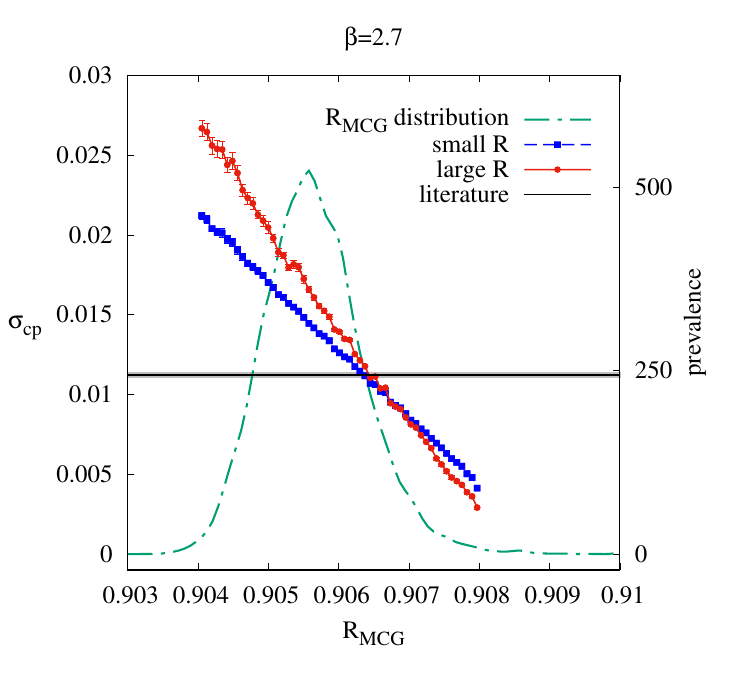}
\caption{Same as Fig.~\ref{fig:bothsigmasRR25} for $\beta=2.7$. But here we observe that in the right tip of the distribution $\sigma_\mathrm{cp}$ deviates drastically from the asymptotic string tension. Further, the slope of $V_\mathrm{cp}(R)$ extracted from small Wilson loops exceeds the slope for large loops.}
\label{fig:bothsigmasRR27}
\end{figure}

\begin{figure}[h!]
\includegraphics[width=.9\linewidth]{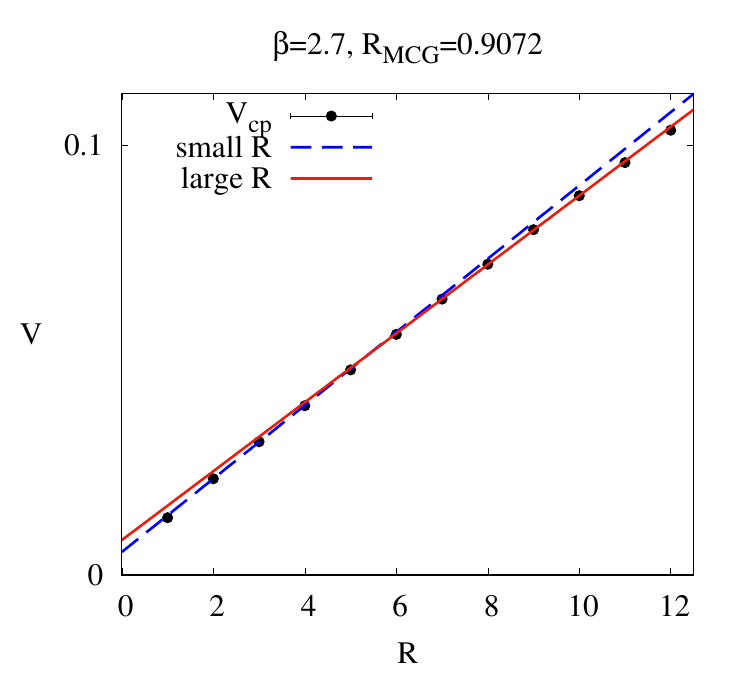}
\caption{Characteristic potential extracted from Wilson loops for $\beta=2.7$ in the large $R_\mathrm{MCG}$ region at $R_\mathrm{MCG}=0.907$. The slope of the center projected potential $V_\mathrm{cp}(R)$ decreases with $R$, as shown by the linear approximations for small and large $R$. This behavior indicates a failure of the vortex detection.}
\label{fig:unusualV}
\end{figure}

Of particular interest is $\beta=2.7$, where we found strong deviations from the Gaussian distribution of $R_\mathrm{MCG}$ values in Fig.~\ref{fig:DistriDistortion}c. The corresponding string tensions are shown in Fig.~\ref{fig:bothsigmasRR27}. 
Note that $\sigma_\mathrm{cp}$ for large R exceeds $\sigma_\mathrm{cp}$ for small R at lower values of $R_\mathrm{MCG}$ whilst the opposite is the case for large values of $R_\mathrm{MCG}$. A potential with such an inverted, wrong behavior of the slope is shown in Fig.~\ref{fig:unusualV} at $R_\mathrm{MCG}=0.907$. Around $R_\mathrm{MCG}=0.908$, we even observe a collapse of the linear behavior and a transition to a screening potential, indicating P-vortices of finite extent.
The unusual, inverted behavior of Fig.~\ref{fig:unusualV} is particularly relevant within the subset of the ensemble that we suggested to discard in the discussion of Fig.~\ref{fig:CDF}. This hints at a possibility to resolve some failures of vortex detection for $\beta=2.7$ where the vortex detection by center vortices became more complicated.

\section{Removal of vortex detection failures}
\begin{figure}[h!]
\includegraphics[width=\linewidth]{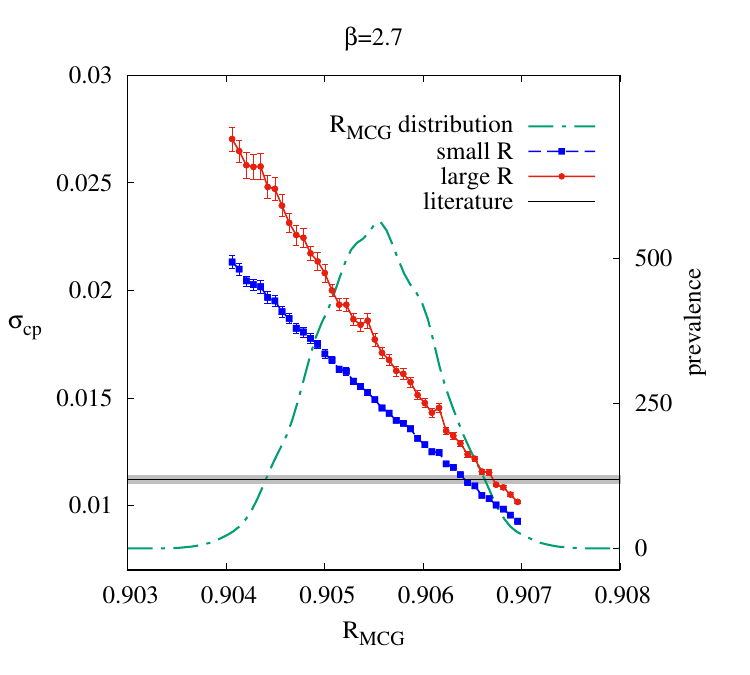}
\caption{Same as Fig.~\ref{fig:bothsigmasRR25} for $\beta=2.7$ after removing from the ensemble Fig.~\ref{fig:DistriDistortion}c configurations where vortex detection appeared to have failed. The discrepancy obviously present in Fig.~\ref{fig:bothsigmasRR27} is weakened: The underestimation of the string tension is improved.}
\label{fig:reasonableRR27}
\end{figure}

For sufficiently large $\beta$, lattices of size $24^4$ are usually large enough to extract potentials from single field configuration by evaluating their Wilson loop data. This observation allows for $\beta=2.7$ to identify local maxima where vortex identification fails. We assume that this failing is caused by vortices being partly overlooked which in turn resembles a broken vortex percolation. From the 20\,000 configurations we have removed those whose string tension decrease with $R$, i.e. with a behavior as shown in Fig.~\ref{fig:unusualV}. This results in a modified, more symmetric distribution of the remaining 16\,000 local maxima of $R_\mathrm{MCG}$, see Fig.~\ref{fig:reasonableRR27}. The string tensions extracted from averages over 200 center projected Wilson loops are shown in Fig.~\ref{fig:reasonableRR27}.
The intersection of $\sigma_\mathrm{cp}$ with the horizontal line has changed only slightly from $R_\mathrm{MCG}=$0.9064 to 0.9066 due to the modification of the ensemble. The agreement between literature value and extracted string tension corresponds to the right tip of the distribution, suggesting that the largest maxima within the symmetrized ensemble of local maxima predict the physical string tension. Hence, we assume that even for $\beta=2.7$, the ensemble of local maxima of the gauge functional contains the information about the density of thick vortices when obvious failures of the vortex identification are discarded from the ensemble.

\section{Towards a modified definition of MCG}
In the previous sections and in Fig.~\ref{fig:DistriDistortion} we have found that the distributions of local maxima are symmetric for $\beta\in\{2.3,2.4,2.5\}$, have a tiny asymmetry for $\beta=2.6$ and a strong asymmetry for $\beta=2.7$. Further, for the 20\,000 fields at $\beta=2.7$, there are several thousands with a slope of $V_\mathrm{cp}(R)$ decreasing with $R$. We remove these fields with wrong behavior of the slope. Then we symmetrise the ensembles for $\beta=2.6$ and $2.7$.

\begin{figure*}
\includegraphics[width=.328\linewidth]{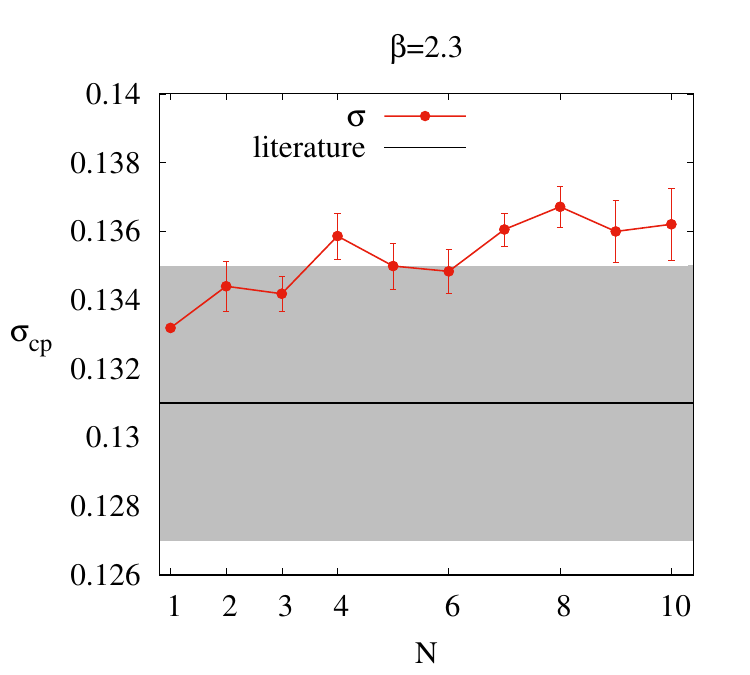}
\includegraphics[width=.328\linewidth]{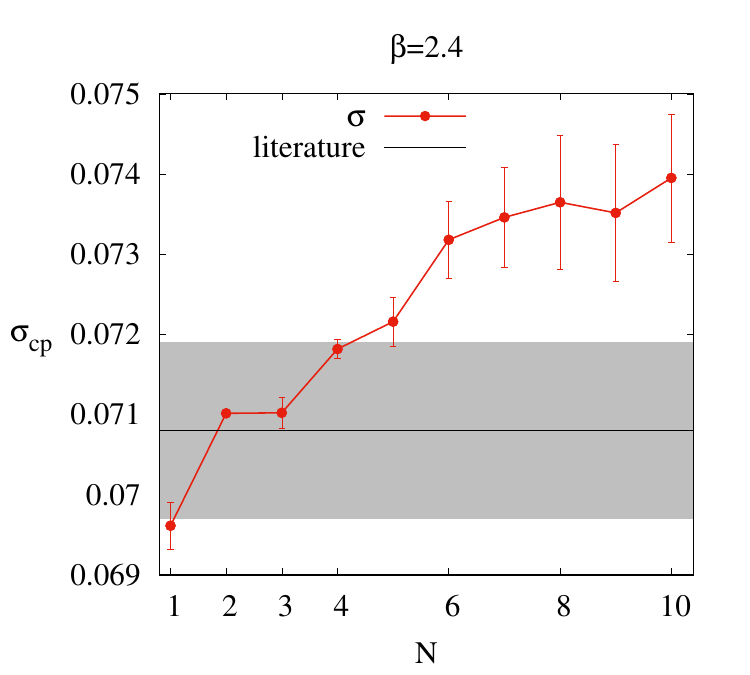}
\includegraphics[width=.328\linewidth]{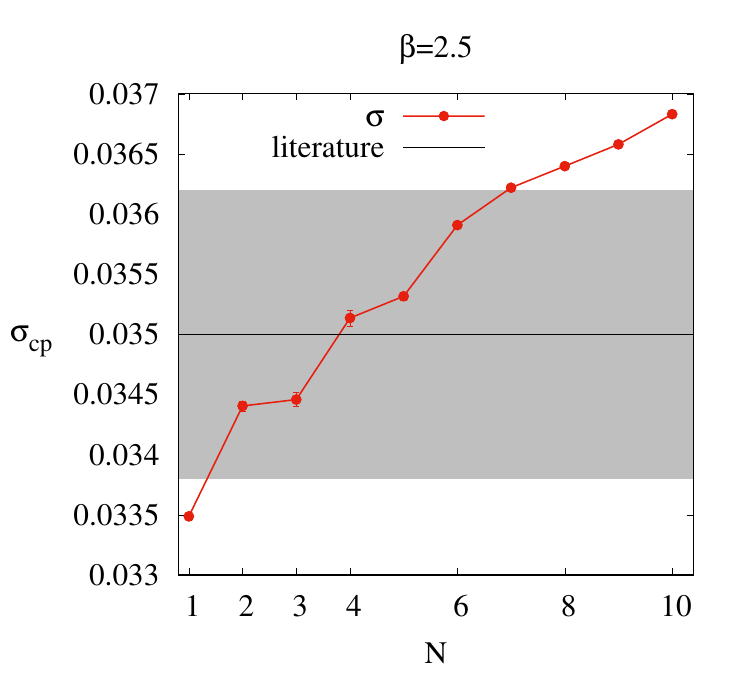}
\\
\includegraphics[width=.328\linewidth]{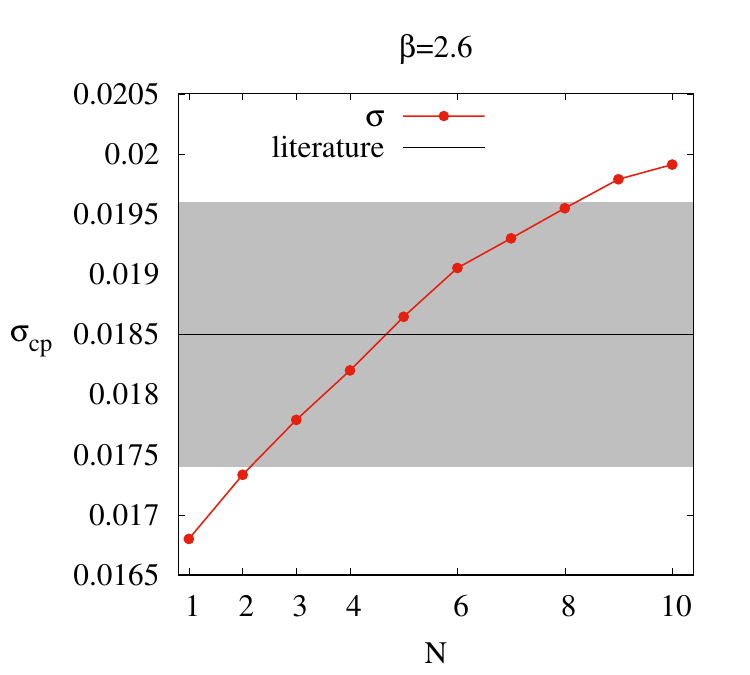}
\includegraphics[width=.328\linewidth]{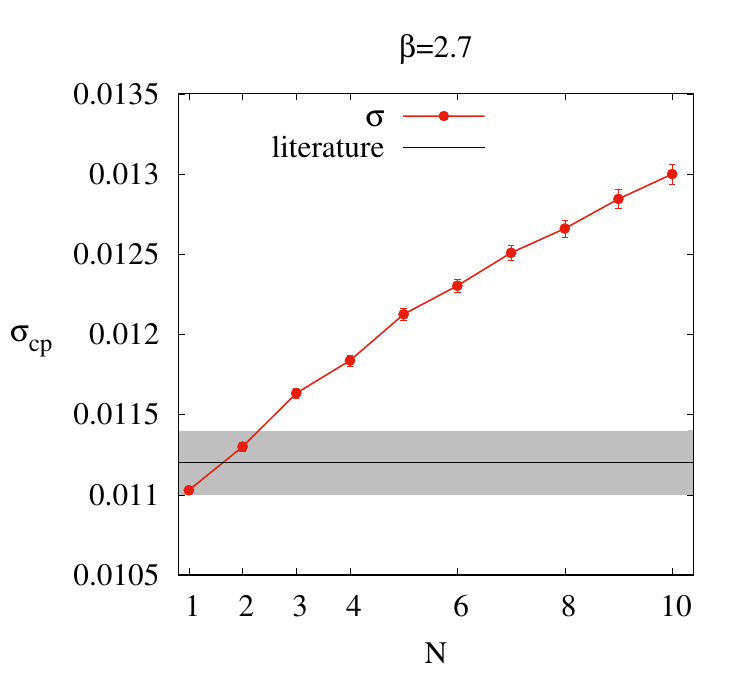}$\quad$
\includegraphics[width=.273\linewidth]{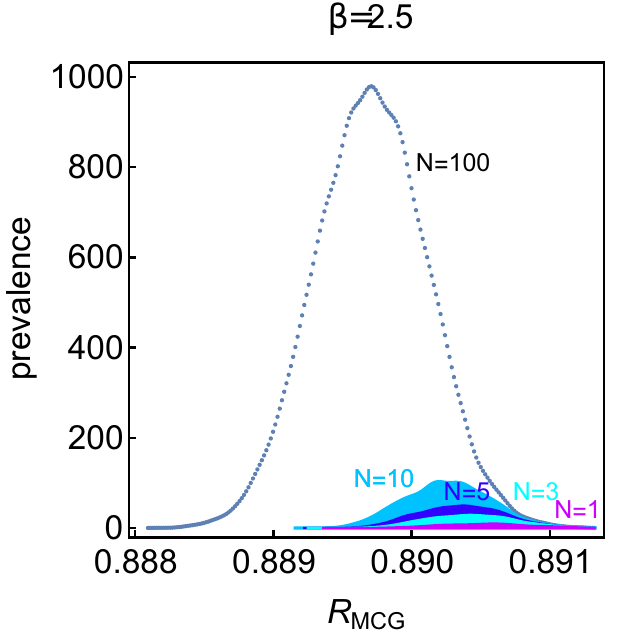}
\caption{\small For $\beta\in\{2.3, 2.4, 2.5, 2.6, 2.7\}$ on a $24^4$ lattice, string tensions are extracted from center projected Wilson loops of 200 different physical fields from the $N$ gauge copies with the highest values of $R_\mathrm{MCG}$. For $\beta=2.6$ a symmetrisation of the distribution removes some of the highest extreme $R_\mathrm{MCG}$ values. At $\beta=2.7$ configurations with P-vortices which did not percolate had to be removed. The horizontal black lines indicate string tensions extracted from unprojected configurations in Refs.~\cite{Caselle:2015tza, Fingberg:1992ju, Bali1994, Michael:1990fh}, $\sigma = 0.131(4), 0.0708(11), 0.0350(12), 0.0185(11), 0.0112(2)$ for $\beta=2.3$ to $2.7$, respectively. The last figure compares the normalized distribution of local maxima of $R_\mathrm{MCG}$ at $\beta=2.5$ for several $N$. We observe that the average of these $R_\mathrm{MCG}$ distributions moves with increasing $N$ to lower values and reaches for $N=100$ the average of the whole ensemble of gauge copies.}
\label{fig:SN}
\end{figure*}

The success of DMCG~\cite{DelDebbio:1998luz,hollwieser:2008tq} with the request of global maximization of the gauge functional, done for a few gauge copies only, leads to the conjecture that the fields with the highest $R_\mathrm{MCG}$ values within the remaining symmetrized ensembles allows to predict the string tensions of the full quark-antiquark potentials. We are now going to check this hypothesis for our ensembles which are symmetrized for $\beta=2.6$ and corrected and symmetrized for $\beta=2.7$. We determine $\sigma_\mathrm{cp}$ from the asymptotic slopes of the center projected quark-antiquark potential. For the 200 physically different fields we select the $N$ gauge copies with the highest values of $R_\mathrm{MCG}$ within the remaining ensembles. Since the averages of the ensembles lead generally to an overestimate of the string tension~\cite{dehghan_2023}, we can expect that the extracted string tension approaches the literature value for decreasing $N$ from above. The asymptotic slopes for various values of $N$ are compared in Fig.~\ref{fig:SN} with the literature values of the string tension. 

We observe that already a few gauge copies, with the highest values of $R_\mathrm{MCG}$, are sufficient to approach the literature values of string tensions extracted from unprojected field configurations. At $\beta=2.3$ the strong coupling is largest. This manifests itself in the highest value of the string tension. The expected values of the Wilson loops therefore disappear most quickly in statistical noise before the asymptotic value of $\sigma_\mathrm{cp}$ is possibly reached. This could be the reason why the string tension for beta=2.3 is already overestimated for $N=1$. Also for $\beta=2.4$ and 2.5 no special action is necessary. For two or three gauge copies we get already agreement between $\sigma_\mathrm{cp}$ for large Wilson loops and the accepted values of the string tension. At $\beta=2.6$ a few gauge copies with extreme values of $R_\mathrm{MCG}$ had to be removed by symmetrisation of the ensemble, as discussed in section four. At $\beta=2.7$ failures of the vortex detection can be identified by missing percolation of vortices leading to a decrease of the string tension with the size of center projected Wilson loops. The remaining symmetrized ensemble allows for the same accuracy in the determination of the asymptotic string tension as for lower $\beta$ values. The last diagram of Fig.~\ref{fig:SN} depicts the distribution of $R_\mathrm{MCG}$ values for various choices of the number $N$ of gauge copies for $\beta=2.5$. These distributions are again nearly Gaussian. With increasing $N$ their average moves towards lower $R_\mathrm{MCG}$ values. Due to the almost linear relation between $R_\mathrm{MCG}$ and $\sigma_\mathrm{cp}$, see Fig.~\ref{fig:bothsigmasRR25}, increasing $N$ results in an increased string tension.

\section{Conclusion}
In this article it was shown that the ensemble of local maxima of the MCG gauge functional $R_\mathrm{MCG}$ contains the information about the long range behavior of the quark-antiquark potential. To counter an underestimation of the string tension the request for global maximization of the gauge functional was substituted by the request of maximization within the symmetric Gaussian ensemble of local maxima. We have shown that there is a window of $\beta$ values where the asymptotic string tension of center projected Wilson loops can be easily determined since the ensemble of $R_\mathrm{MCG}$ values is well Gaussian distributed. By restricting the maximization to this Gaussian subset of the ensemble extreme values of the gauge functional do not play any role due to their small weights. The tip of maximal $R_\mathrm{MCG}$ values within these Gaussian distributions determines the gauge copies for which the center projected string tensions $\sigma_\mathrm{cp}$ agree well with the expected values of the asymptotic string tension for unprojected fields.

The $\beta$-window can be enlarged towards higher values by symmetrization of the ensembles and diminishing the weight of large $R_\mathrm{MCG}$ values. A further increase of the window is possible by excluding gauge copies experiencing an inverted behavior of the slope of the potential with respect to the loop sizes. Due to finite separation and thickness of vortices on the lattice, one can only expect limited increase of the $\beta$-window. Within this window the analysis of the ensemble shows that the center vortex model is capable to reproduce the correct string tension by center projection.

An attempt to systematically address the issue of a heuristically defined gauge has been presented in the context of Coulomb gauge~\cite{Heinzl_2008}.

\bibliography{refs}

\end{document}